\documentclass[authoryear,preprint,10pt,review]{elsarticle}
\usepackage{amssymb,amsmath,gensymb,multirow}
\usepackage{lineno}
\usepackage{float}
\usepackage{color}
\journal{arXive.org} 
\begin{document}
\begin{frontmatter}
\title{Mean and variance of first passage time in Markov chains with unknown parameters}

\author[UC,SL]{Carlos Hernandez-Suarez\corref{cor1}}
 \ead{carlosmh@mac.com}

\address[UC]{Facultad de Ciencias, Universidad de Colima, Bernal Diaz del Castillo 340, Colima, Colima, 28040, MEXICO}
  \address[SL]{Simon A. Levin Mathematical and Computational Modeling Sciences Center, Arizona State University, Tempe, AZ 85287-3901}

 \cortext[cor1]{Corresponding author}

\begin{abstract}

There are known expressions to calculate the moments of the first passage time in Markov chains. Nevertheless, it is commonly forgotten that in most
applications the parameters of the Markov chain are constructed using estimates 
based upon empirical data and in those cases the data sample size should play an important role in estimating the variance.
 Here we provide a Monte Carlo approach to estimate the first two moments of the passage time in this situation. We illustrate this method with an example using data from the biological field.
\end{abstract}


\end{frontmatter}

\section{Introduction}
 In an ergodic Markov chain with $m$ states, \emph{first passage time} is defined as the time to reach a particular set of states $S$ for the first time, starting from a state distribution $\mathbf{v}=[ v_1,v_2,\ldots,v_k], k <m$. The first passage time is important, for instance, in Markov Population Models  to calculate the longevity of individuals as well as other relevant life history traits (LHTs), like generation time and the basic reproductive number \citep{caswell2009stage}. 
 
 In very few real life situations, the parameters of a Markov chain are known exactly, and when this occurs, is due mostly by model assumptions. For instance, consider a Markov chain with states corresponding to the outcome when rolling a dice, thus $W=\{1,2,3,4,5,6\}$ and we assume the transition probabilities between states $i$ and $j$ as $p_{ij}=1/6$ by model, but most of the time, the parameters of the Markov Chain have inherent error, which is closely related to the sample size. 
 
 We can use $n_{ij}/n_i$, the fraction of transitions from state $i$ to $j$ out of $n_i$ as an estimate of $p_{ij}$,  \citep{GuttorpPeter1995Smos}.  We provide a simple methodology based in Monte Carlo simulations that incorporates sample size in our estimation of the first passage time in Markov chains.
 
 \section{Methodology}
 
To calculate the fist passage time from an initial state $i_0$ to a set of states $S$, we can transform the Markov chain so that every state in $S$ is absorbing and then analyze the time to extinction. Assume a Markov chain has $k$ transient and $r$ absorbing states that is decomposed as:

\[
\mathbf{P}=
\begin{pmatrix}
\mathbf{U} & \mathbf{R} \\
\mathbf{Z} & \mathbf{I} 
\end{pmatrix}
\]
\noindent
 where $\mathbf{U}$ is a $k \times k$ matrix containing the transitions between the transient states, $\mathbf{R}$ is a $k \times r$ matrix containing the transitions between transient states to absorbing states, $\mathbf{Z}$ is a $r \times k$ matrix of zeros and  $\mathbf{I}$ is the identity matrix of size $r$. Then, starting from the initial distribution $\mathbf{v}'=[ v_1,v_2,\ldots,v_k]$, the expected value and variance of the time to extinction of the process, $L$, are:
 
\begin{eqnarray}
E[L] &=&\mathbf{\underline{v}}' \mathbf{N} \mathbf{\underline{1}}\nonumber\\
V[L]&=&  \mathbf{\underline{v}}'\mathbf{N} (2\mathbf{N} -\mathbf{I})\mathbf{\underline{1}}\ -E[L ] E[L ]
\label{eq1}
\end{eqnarray}
 \noindent
 where $\mathbf{1}$ is a column vector of 1's and $\mathbf{N}=( \mathbf{I}- \mathbf{U})^{-1}$ is the \emph{fundamental matrix}  \citep{iosifescu1980finite}.
 
 If $\mathbf{U}$ is unknown, its parameters are estimated through observation or experimentation, giving rise to a matrix $\mathbf{\hat{U}}$, some random realization of matrix $\mathbf{U}$, thus, in the most common scenario, the moments in (\ref{eq1}) are not calculated using $\mathbf{U}$ but instead, with a particular realization $\mathbf{\hat{U}}$.
 
Let $\mathbf{\hat{N}}=( \mathbf{I}- \mathbf{\hat{U}})^{-1}$ be an estimate of the fundamental matrix $\mathbf{N}$. The Variance of the passage time can be written by conditioning on $\mathbf{\hat{U}}$ :
 
\begin{eqnarray}
V[L] &= &E[V[L | \mathbf{\hat{U}}] + V[E[L | \mathbf{\hat{U}}]\nonumber\\
&= &E[( \mathbf{v}'\mathbf{\hat{N}} (2\mathbf{\hat{N}} -\mathbf{I})\mathbf{1} -E[\mathbf{v}' \mathbf{\hat{N}} \mathbf{1} ] E[\mathbf{v}' \mathbf{\hat{N}} \mathbf{1}]] + V[\mathbf{v}' \mathbf{\hat{N}} \mathbf{1}] 
\label{eq2}
\end{eqnarray}
 \noindent
Row $i$ of $\mathbf{U}$ has parameters $\mathbf{\underline{u}}'=[p_{i1},p_{i2},\ldots, p_{ik}]$ that were estimated with $\mathbf{\underline{\hat{u}}}'=[x_{i1},x_{i2},\ldots, x_{ik}]n_i^{-1}$, the observed fraction of transitions from state $i$ to each state. A sample of size $n_i$ from $\mathbf{\underline{u}}$  follows a multinomial distribution with parameters $\mathbf{\underline{u}}$ and $n_i$, with expected value $\mathbf{\underline{u}}$ and variance-covariance matrix $\Sigma$:

\[
\mathbf{\underline{u}}=
n_i \begin{pmatrix}
\hat{p}_{i1}\\
\hat{p}_{i2}\\
\vdots \\
\hat{p}_{ik}
\end{pmatrix}\ \ \ \
\mathbf{\Sigma}= 
n_i^{-1}\begin{pmatrix}
\hat{p_{i1}} (1-\hat{p}_{i1}) & -\hat{p}_{i1} \hat{p}_{i2}  & \cdots & -\hat{p}_{i1} \hat{p}_{ik} \\
-\hat{p}_{i1} \hat{p}_{i2}  & \hat{p_{i2}} (1-\hat{p}_{i2})  & \cdots & -\hat{p_{i2}} \hat{p}_{ik}  \\
\vdots & \vdots & \ddots & \vdots\\
 -\hat{p}_{i1} \hat{p}_{ik}  & \hat{p_{i2}} \hat{p}_{ik}   & \cdots & \hat{p}_{ik} (1-\hat{p}_{ik})
\end{pmatrix}\  
\]

\noindent
For details see  \cite{GuttorpPeter1995Smos}, Theorem 2.16, p.65 

We can sample from the multinomial distribution for each row of $\mathbf{U}$, that is, simulate $\mathbf{\hat{U}}$ and use it to calculate $\mathbf{\hat{N}}$ and thus the moments in (\ref{eq1}). The average and variance over the simulations can be plugged in (\ref{eq2}) to provide an estimate of the variance of the passage time.

We can thus summarize the Monte Carlo estimation as follows:

\begin{enumerate}
\item For every row $\underline{\hat{\mathbf{u}}_i}$ in $\mathbf{\hat{U}}$, simulate a vector of transitions $\underline{\mathbf{x}_i}$ from a multinomial distribution with parameters $n_i$ and $\underline{\hat{\mathbf{u}}_i}$, where $n_i$ is the total number of transitions observed from state $i$ to all other states. Let $\underline{\hat{\mathbf{p}}_i}= \underline{\hat{\mathbf{u}}_i}/n_i$
\item Use the samples of the previous step to construct a matix $\mathbf{\hat{U}^*}$.
\item Use the matrix $\mathbf{\hat{U}^*}$ to calculate  $\mathbf{N^*}=( \mathbf{I}-\mathbf{\hat{U}^*} )^{-1}$ and with (1) calculate $\mu_j =E[L]$ and $\sigma_j=V[L]$. Record these $\mu_j$ and  $\sigma_j$ as the outcome of the $j$-th simulation. 
\item Repeat from step 1.
\end{enumerate}

The average of the $\sigma_i$'s is used to estimate $E[V[L | \mathbf{\hat{U}}]]$, while the variance of the $\mu_j$'s estimates $V[E[L | \mathbf{\hat{U}}]]$. Adding these two yields an estimate of the variance of the Longevity, as given in (\ref{eq2}).

\section{Example}

Biologists use Markov chains to model the transition of individuals through different developmental stages. Death is an absorbing state. Due to the use of the recurrence

$$
\mathbf{\underline{X}} (t+1) = \mathbf{A} \mathbf{\underline{X}} (t)
$$
to express the change in the stage-specific composition of the population at two subsequent discrete generations, it is customary among biologists to express Markov
chains using its transpose,  which is called a Matrix Population Model, MPM, for details, see \cite{caswell2009stage}, \cite{hernandez2019building}.

\begin{figure}[ht]
\begin{center}
\includegraphics[width=4in]{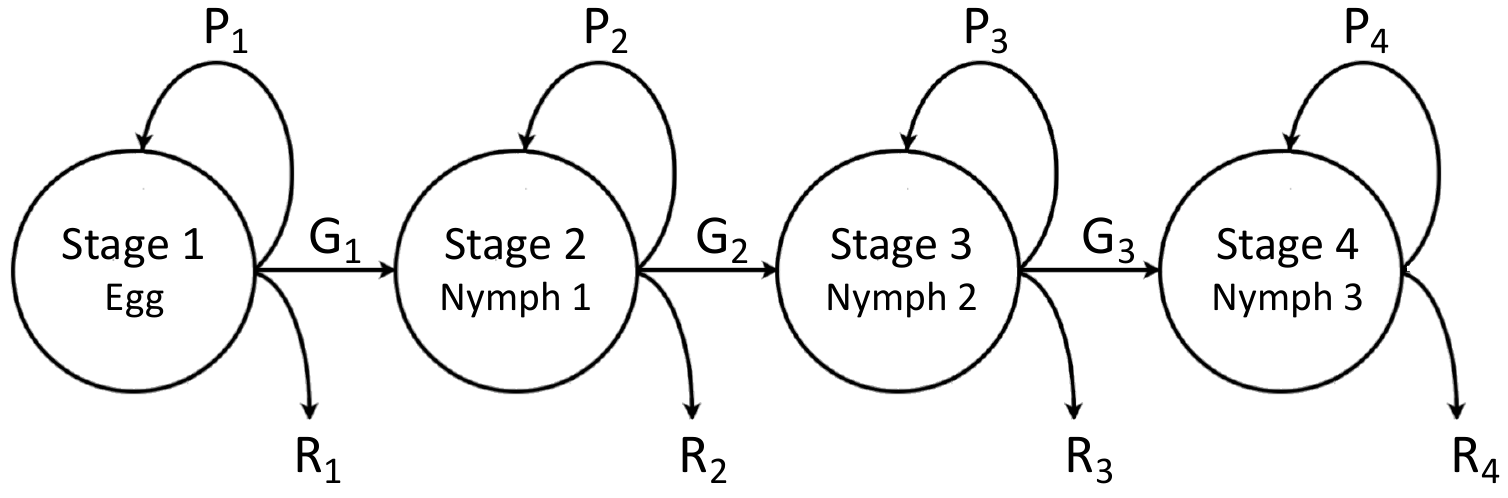}
\caption{An example life cycle graph with five stages.}
\label{Fig1}
\end{center}
\end{figure}

An example of transitions between four stages is depicted in Figure (\ref{Fig1}). In a given state $i$, an individual can move to the next stage with probability $G_i$ (graduation), stay one more unit of time in that state with probability $P_i$ or die with probability $R_i$. 

In \cite{hernandez2019building}, a data set for the life cycle of the kissing bug \emph{Eratyrus mucronatus} (Reduviidae, Triatominae) was introduced. There are seven stages: egg, nymph states 1-5 and the adult stage. Table (\ref{TableGyP}) shows the estimates of each of the parameters and the sample size to estimate the parameters of each stage. For details in the estimation method, see \cite{hernandez2019building}.

 \begin{table}[h]
\caption{Estimates of the stage-specific parameters of a matrix population model for the kissing bug \textit{Eratyrus mucronatus}, from non-individualized cohorts reared in the laboratory at $16 ^\circ$C and 65\% relative humidity (see suppl. material in \cite{hernandez2019building}).}
\begin{center}
\begin{tabular}{ccccc}
\hline
Stage & $\hat{G_i}$  & $\hat{R_i}$ & $\hat{P_i}$ & $n^*$ \\
\hline
Egg    &    139/676    &    59/676    &    478/676  & 676   \\
N1    &    89/669    &    52/669    &   528/669  & 669    \\
N2    &    74/390    &    15/390    &    301/390  & 390    \\
N3    &    60/466    &    14/466    &    392/466  & 466    \\
N4    &    59/465    &    1/465    &    405/465 & 465    \\
N5    &    55/912    &    4/912    &    853/912  &  912    \\
Adult    &    0    &    55/2570    &    2515/2570  & 2570    \\
\hline
\multicolumn{5}{c}{(*) Total number of transitions observed per stage.} \\
\end{tabular}
\end{center}
\label{TableGyP}
\end{table}
 
 The estimated matrix $\mathbf{\hat{U}}$ for the (transposed) Markov Chain is:
 
 \begin{equation*}
\mathbf{\hat{U}}=
\begin{pmatrix}
478/676 & 0 & 0 & 0 & 0 & 0 & 0\\
139/676 & 528/669 & 0 & 0 & 0 & 0 & 0 \\
 0 & 89/669 & 301/390 & 0 & 0 & 0 & 0 \\
 0 & 0 & 74/390 & 392/466 & 0 & 0 & 0\\
 0 & 0 & 0 & 60/466& 405/465 & 0 & 0\\
 0 & 0 & 0 & 0 & 59/465 & 853/912 & 0\\
 0 & 0 & 0 & 0 & 0 & 55/912 & 2515/2570
\end{pmatrix}\ \ \ \
\label{Uest}
\end{equation*}
\noindent

\subsection{Monte Carlo estimation}

For every stage $i$, the parameter space has expected value $\mathbf{\underline{u}}$ and variance-covariance matrix $\Sigma$:

\begin{equation}
\mathbf{\underline{u}}=
\begin{pmatrix}
\hat{G}_i\\
\hat{P}_i\\
\hat{R}_i
\end{pmatrix}\ \ \ \
\mathbf{\Sigma}= 
n_i^{-1}\begin{pmatrix}
\hat{G_i} (1-\hat{G}_i) & -\hat{G}_i \hat{P}_i & -\hat{G}_i \hat{R}_i \\
-\hat{P}_i \hat{G}_i  & \hat{P_i} (1-\hat{P}_i)   & -\hat{P_i} \hat{R}_i   \\
\hat{R}_i \hat{G}_i  & -\hat{R_i} \hat{P}_i    & \hat{R}_i (1-\hat{R}_i)
\end{pmatrix}\  
\label{varcov}
\end{equation}

 Using this data, we performed $1 \times 10^5$ simulations and obtained the sample mean and variance and the moments obtained using (\ref{eq1}). These are shown in Table \ref{comparison}.
 
\begin{figure}[ht]
\begin{center}
\includegraphics[width=4in]{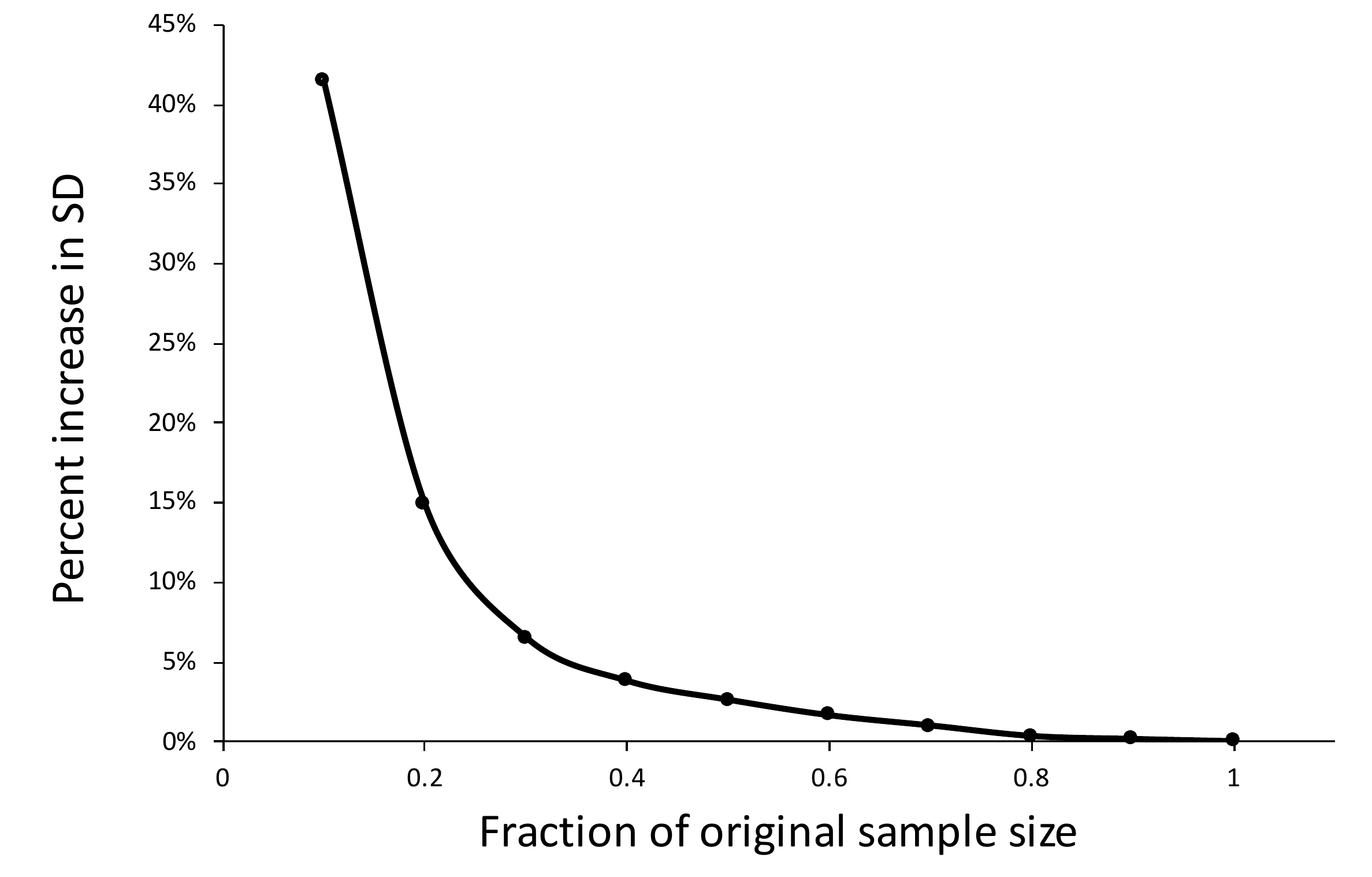}
\caption{Effect of sample size on the standard deviation of the longevity. The percentage increase at different fractions of the original sample size are shown.}
\label{Fig2}
\end{center}
\end{figure}

 We analyzed the effect of the sample size on the variance of the longevity. To do this, we reduced the number of transitions (last column of Table \ref{TableGyP})  to a fraction $f$ of the original size, and use these for the simulation process. The effect of the reduction in sample size is shown in Figure \ref{Fig2}.
 
 \section{Discussion}
 The difference between the calculated expected values and standard deviation of the longevity using (\ref{eq1}) vs. those obtained using Monte Carlo methods are negligible, which is mainly due to the relatively large sample sizes for every stage in the study (see Table \ref{TableGyP}). In experimental biology, sample sizes may be large for most plants and invertebrates, but is more difficult for vertebrates or even mammals, where follow up is expensive or extremely difficult. Figure \ref{Fig2} shows that when the sample size was reduced to a tenth, the difference between both calculations becomes noticeable: there is an increase of $41\%$ in the standard deviation between both Monte Carlo simulations.  
 
 It must be noticed that in the simulations with sample size reduced, we only reduced the value of $n_i$ in eq. (\ref{varcov}) but kept the same estimates of $P_i$, $G_i$ and $R_i$ shown in Table {\ref{TableGyP}. Since these values were achieved with the original sample size, we expect that the increase in variance caused by a reduction in sample size would be larger than the indicated in Fig. \ref{Fig2}.

 \begin{table}[H]
\caption{Comparison of expected value and standard deviation for the longevity (in weeks) of the kissing bug \textit{Eratyrus mucronatus}. Results of $1 \times 10^5$ simulations.}
\begin{center}
\begin{tabular}{lcc}
 & Calculated$^\dagger$  & Monte Carlo  \\
\hline
Expected value    &    30.658    &    31.087    \\
Standard deviation    &    44.943    &    46.023      \\
\hline
\multicolumn{3}{l}{$^\dagger$ Using equation (\ref{eq1})}\\
\end{tabular}
\end{center}
\label{comparison}
\end{table}

 \section{Supplementary material}
 
 Python code to estimate mean and variance of transition time using multinomial distribution. (Performs $1 \times 10^6$ simulations in about $10$ seconds with a $2.3$ GHz Intel Core i$5$ processor)\\

 \begin{verbatim}
 #!/usr/bin/env python3.6
# -*- coding: utf-8 -*-
import numpy as np

#================ BEGIN DECLARATIONS ==================================
    
#Declaring number of simulations:

nsim = 1000
    
#Declaring sample size for every state:

n=[676,669,390,466,465,912,2570]
#Declaring matrix U:

U=[[478/676,	139/676,	0,	0,	0,	0,	0],
[0,	528/669,	89/669,	0,	0,	0,	0],
[0,	0,	301/390,	74/390,	0,	0,	0],
[0,	0,	0,	392/466,	60/466,	0,	0],
[0,	0,	0,	0,	405/465,	59/465,	0],
[0,	0,	0,	0,	0,	853/912,	55/912],
[0,	0,	0,	0,	0,	0,	2515/2570] ]

#Declaring initial distribution: (start in first stage in this example)

v= np.array([np.zeros(len(U))])
v[0][0] = 1   

#================ END DECLARATIONS ==================================

def simul(U,v,u,I):
    N = np.linalg.inv(I-U)
    p1 = np.matmul(v,N)
    mu = np.matmul(p1,u).tolist()
    mu=mu[0][0] 
    p1 = np.matmul(v,N)
    p2 = np.matmul(2*N-I,u)
    p3 = np.matmul(p1,p2)
    sigma = p3-mu*mu
    sigma = sigma[0][0]
    return mu,sigma
print(' ')
print('Sample sizes =>',n)
I = np.identity(len(U))
u= np.array([np.transpose(np.ones(len(U)))])
u=np.transpose(u)
print(' ')
print('Matrix U:')
print(' ')
for i in U:
    print(i)
    print('sum:',sum(i))
print(' ')
MU = []
SIGMA=[]

count=0
for s in range(nsim):
    Uest = []
    row = 0
    for i in U:
        p=list(i)
        p.append(1-sum(i))
        x = (np.random.multinomial(n[row], p, size=1)/n[row]).tolist()
        Uest.append(x[0][0:-1])
        row += 1
    try:
        mu,sigma = simul(Uest,v,u,I)
        MU.append(mu)
        SIGMA.append(sigma)
    except:
        pass
avsigma = np.mean(SIGMA)
varmu = np.var(MU)
EL = np.mean(MU)
VL = avsigma + varmu
print('RESULTS:')
print('---------------------------------')
print('Expectation => ',EL)
print('Variance    => ',VL)
print('Standar dev.=> ',np.sqrt(VL))
print(' ')
print('Comparison: (no simulations, only expression (1) in draft ')
mu,sigma = simul(U,v,u,I)
print('---------------------------------')
print('Expectation => ',mu)
print('Variance    => ',sigma)
print('Standar dev.=> ',np.sqrt(sigma))
\end{verbatim}

\end{document}